\documentclass[12pt]{spieman}  % 12pt font required by SPIE;
\usepackage{amsmath,amsfonts,amssymb}
\usepackage{graphicx}
\usepackage{setspace}
\usepackage{tocloft}
\usepackage{lineno}
\usepackage{threeparttable}
\nolinenumbers
\title{
In-orbit Demonstration of X-ray Pulsar Navigation with NinjaSat
}

\author[ac*]{Naoyuki Ota}
\author[abc]{Takuya Takahashi}
\author[bac]{Toru Tamagawa}
\author[d]{Tomoshi Takeda}
\author[ef]{Teruaki Enoto}
\author[b]{Takao Kitaguchi}
\author[g]{Wataru Iwakiri}
\author[b]{Yo Kato}
\author[h]{Masaki Numazawa}
\author[ba]{Tatehiro Mihara}
\author[d]{Hiromitsu Takahashi}
\author[i]{Chin-Ping Hu}
\author[ca]{Yuanhui Zhou}
\author[ca]{Keisuke Uchiyama}
\author[ca]{Yuto Yoshida}
\author[ca]{Syoki Hayashi}
\author[ca]{Arata Jujo}
\author[ca]{Sota Watanabe}
\author[cb]{Amira Aoyama}
\author[cb]{Satoko Iwata}
\author[ca]{Kaede Yamasaki}
\author[ca]{Soma Tsuchiya}
\author[ca]{Yosuke Nakano}
\author[g]{Takayuki Kita}
\author[k]{Mayu Ichibakase}
\author[ja]{Hiroki Sato}
\author[l]{Hirokazu Odaka}
\author[m]{Tsubasa Tamba}
\author[b]{Kentaro Taniguchi}

\affil[a]{RIKEN Nishina Center, 2-1 Hirosawa, Wako, Saitama 351-0198, Japan}
\affil[b]{RIKEN Pioneering Research Institute, 2-1 Hirosawa, Wako, Saitama 351-0198, Japan}
\affil[c]{Tokyo University of Science, 1-3 Kagurazaka, Shinjuku, Tokyo 162-8601, Japan}
\affil[d]{Graduate School of Advanced Science and Engineering, Hiroshima University, 1-3-1 Kagamiyama, Higashi-Hiroshima, Hiroshima 739-8526, Japan}
\affil[e]{Department of Physics, Kyoto University, Kitashirakawa Oiwake, Sakyo, Kyoto, Kyoto 606-8502, Japan}
\affil[f]{RIKEN Center for Advanced Photonics, 2-1 Hirosawa, Wako, Saitama 351-0198, Japan}
\affil[g]{International Center for Hadron Astrophysics, Chiba University, 1-33 Yayoi, Inage, Chiba, Chiba 263-8522, Japan}
\affil[h]{Department of Physics, Tokyo Metropolitan University, 1-1 Minamiosawa, Hachioji, Tokyo 192-0397, Japan}
\affil[i]{Department of Physics, National Changhua University of Education, No.1, Jin-De Road,Changhua City, 50007, Taiwan}
\affil[j]{Shibaura Institute of Technology, 307 Fukasaku, Minuma-ku, Saitama 337-8570, Japan}
\affil[k]{Rikkyo University, 3-34-1 Nishi-Ikebukuro, Toshima-ku, Tokyo 171-8501, Japan}
\affil[l]{Department of Earth and Space Science, The University of Osaka, 1-1 Machikaneyama, Toyonaka, Osaka 560-0043, Japan}
\affil[m]{Institute of Space and Astronautical Science, JAXA, 3-1-1 Yoshinodai, Chuo, Sagamihara, Kanagawa 252-5210, Japan}

\usepackage{xcolor}
\newcommand{\rev}[1]{\textcolor{black}{#1}}
\newcommand{\engrev}[1]{\textcolor{black}{#1}}

\cftpagenumbersoff{figure}
\cftpagenumbersoff{table} 
\begin{document} 
\maketitle

\begin{abstract}
\engrev{This study} demonstrated the pulsar navigation capability of a CubeSat X-ray observatory, NinjaSat, equipped with two sets of Gas Multiplier Counters (GMCs).
GMCs are sensitive to \engrev{the} 2--50~keV band with an effective area of 16 cm$^2$ per module at 6~keV.
We verified the timing accuracy by observing  Crab Pulsar and confirmed \engrev{a stable} timing measurement performance is stable within 100~$\mu$s.
To demonstrate the pulsar navigation, we applied a method that optimizes orbital parameters to maximize the significance of the pulse profile of X-rays from a pulsar \engrev{using} "Significance Enhancement of Pulse-profile with Orbit-dynamics \engrev{(SEPO)}". 
We observed Crab Pulsar with an exposure time of \engrev{approximately} 100~ks at different epochs and analyzed the data downlinked to the ground.
By comparing the optimized orbit with the satellite position derived from Global Positioning System data, we quantitatively evaluated the pulsar navigation performance. 
The results showed that the position component along the Crab line of sight was consistently constrained within \rev{$\sim$40~km}, \engrev{and} the range of three-dimensional error varied between \rev{27--370~km} depending on the observation epoch. 
This study demonstrates the feasibility of applying a CubeSat-class X-ray observatory to pulsar navigation, and also provides the first experimental verification that the accuracy of the SEPO method depends on the seasonal geometry between the orbital plane and the pulsar direction.

\end{abstract}


{\noindent \footnotesize\textbf{*}Address all correspondence to Naoyuki Ota,  \linkable{naoyuki.ota@a.riken.jp} }

%\begin{spacing}{2}  
\begin{spacing}{1}  

\section{Introduction}
\label{sec:intro}  
In deep space exploration, autonomous navigation systems are essential because \engrev{Global Navigation Satellite Systems (GNSSs)} cannot provide coverage beyond Earth orbit.
X-ray pulsar navigation has been proposed as \engrev{a} technique for deep-space autonomous navigation.~\cite{kohlhase1975autonomous}~\cite{chester1981navigation}~\cite{WANG202344} 
This method estimates a spacecraft position using periodic signals emitted by neutron stars known as X-ray pulsars.
\engrev{Pulsars are rapidly spinning neutron stars, emitting X-rays via synchrotron radiation from relativistic charged particles.}
When observed from a sufficient distance, these emissions appear as periodic pulses, with highly stable periods corresponding to a spin of a pulsar. 
By accurately measuring the arrival times of these pulses and comparing them with predicted pulse phases based on a known timing model, it is possible to estimate the location of the observatory.

\engrev{This concept can be implemented using the “Significance Enhancement of Pulse-profile with Orbit-dynamics” (SEPO) method, which searches for the orbital parameters that maximize the coherence of the pulsed signal from a single pulsar at the start of the observation.}
\engrev{This method was demonstrated at low Earth orbits (LEOs) by the large-scale X-ray observatory Insight-HXMT, as reported by Zheng et al.~\cite{Zheng_2019}, and a large-scale gamma-ray polarimeter POLAR, as reported by Zheng et al. ~\cite{polar}.}
\engrev{The Grouping bi-$\chi^2$ method is an extension of the SEPO method validated using Rossi X-ray Timing Explorer (RXTE) observational data~\cite{SUN2023386}.}
The method of tracking pulsar phases and providing one-dimensional positioning along the line of sight to each pulsar has been demonstrated by the Station Explorer for X-ray Timing and Navigation Technology of the Neutron-star Interior Composition Explorer (NICER)~\cite{10.1117/12.2231304},\engrev {~\cite{mitchell2018sextant} and X-ray Pulsar Navigation-I mission~\cite{xpnav-1}.}
However, in deep space exploration, onboard resources, such as instrument size and power consumption, are limited. Therefore, X-ray detectors used for pulsar navigation must be compact, lightweight, and operational with \engrev{low-power} consumption.
\engrev{However, the effective area of small instruments is limited, and the number of observable pulsars is restricted. 
Therefore, this study adopted the SEPO method, which performs navigation using a single pulsar.}

\begin{figure} [ht]
\begin{center}
%\begin{tabular}{c} 
\includegraphics[width=1\linewidth]{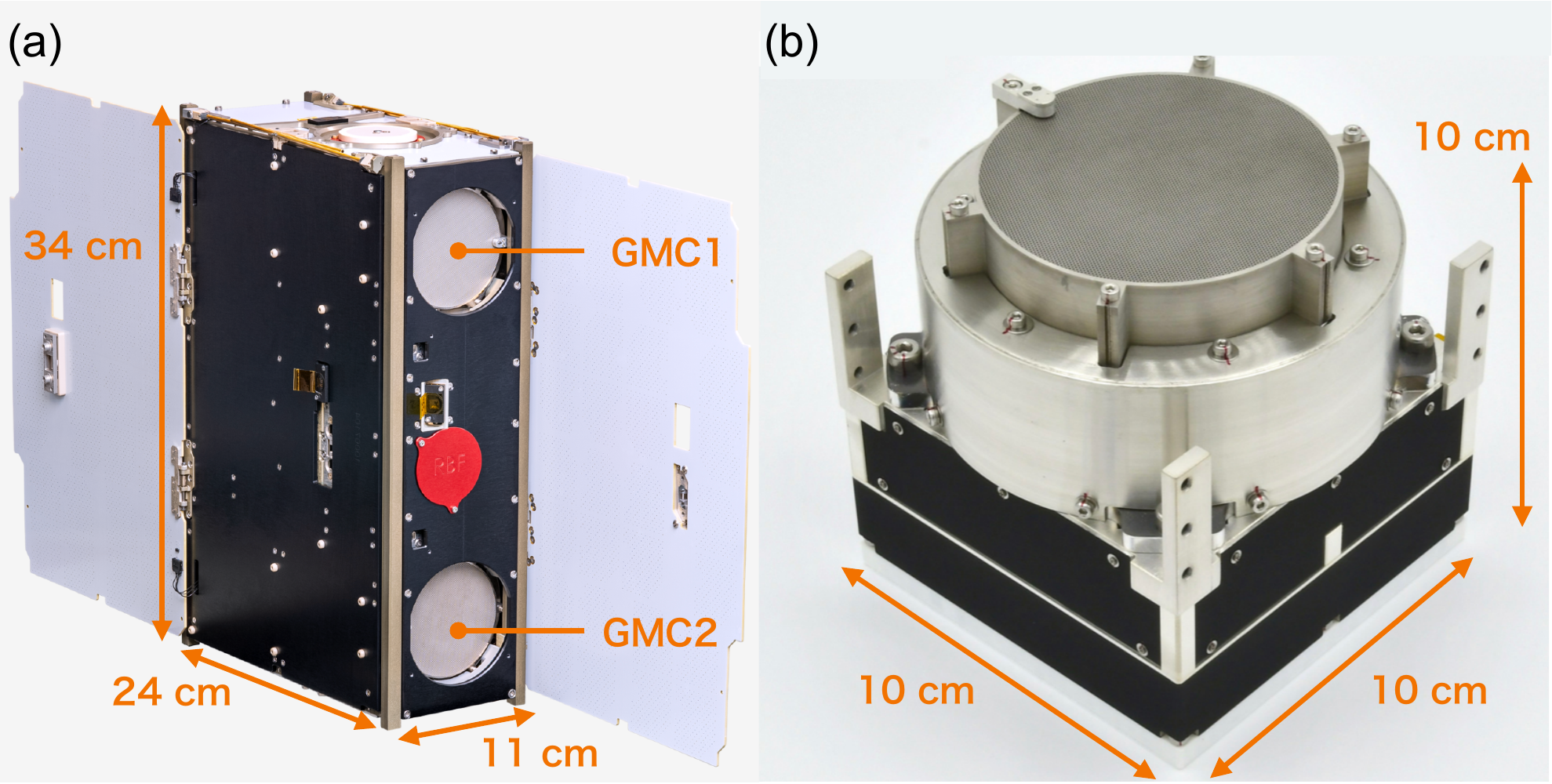}
%\end{tabular}
\end{center}
\caption[example] 
{(a) A photograph of NinjaSat. 
(b) A photograph of the GMC.  
\label{fig:OverviewPhoto} }
\end{figure}

\engrev{NinjaSat is a CubeSat X-ray observatory\cite{10.1093/pasj/psaf014}, whose 
mission concept is to perform long-term monitoring of X-ray sources, such as black holes and neutron stars, and rapid follow-up observation of transient events.
Moreover, it provides an opportunity to demonstrate pulsar navigation with compact instrumentation.
The size of NinjaSat is 6U (11$\times$24$\times$34~cm$^3$), as shown in Fig.~\ref{fig:OverviewPhoto}(a).}
NinjaSat was deployed into a sun-synchronous polar orbit (SSO) at an altitude of 530~km and Local Time of Descending Node of 10:32~a.m.
Its total mass is 8~kg, and the maximum power consumption is 16~W. 
The primary science payloads on NinjaSat are two sets of  Gas Multiplier Counters (GMCs).
They are 1 U (10$\times$10$\times$10~cm$^3$) size detectors\engrev{,} as shown in Fig.~\ref{fig:OverviewPhoto}(b).
The GMC is sensitive to the 2--50~keV band with an effective area of 16~cm$^2$ at 6~keV.
The weight of the GMC is 1.2~kg, and the maximum power consumption is 2.0~W. 

\engrev{NinjaSat is equipped with a flight computer (FC), payload controller (PC), and Global Positioning System (GPS) receiver.
The GPS receiver is compatible with GNSS. However, in accordance with NinjaSat conventions, the term “GPS” is used to refer to GNSS.}
The GPS receiver provides absolute time and position data.
\engrev{A payload controller receives data from GMCs and records it.}
FC includes an Attitude Determination and Control Subsystem (ADCS).
\engrev{It maintains pointing accuracy and a stability of 0.1$^{\circ}$ relative to a target source in 2 $\sigma$.  
In this study, we extended the application of pulsar navigation to a CubeSat-class platform, demonstrating its feasibility with the CubeSat X-ray observatory, NinjaSat, at LEO.}
In Section~\ref{sec:InstAndCal}, we describe the timing system of NinjaSat and its verification results.
In Section~\ref{sec:xnav_method}, \ref{sec:ObsAndData}, \ref{sec:xnav_results}, and \ref{sec:discussion}, we present the method, data, results and discussion of the in-orbit demonstration of pulsar navigation.

%\section{Instruments and Calibration}
\section{Timing Calibration}

\label{sec:InstAndCal}
\subsection{The Timing Systems of NinjaSat}
\label{subsec:timing_system}
\begin{figure} [ht]
\begin{center}
\includegraphics[width=0.9\linewidth]{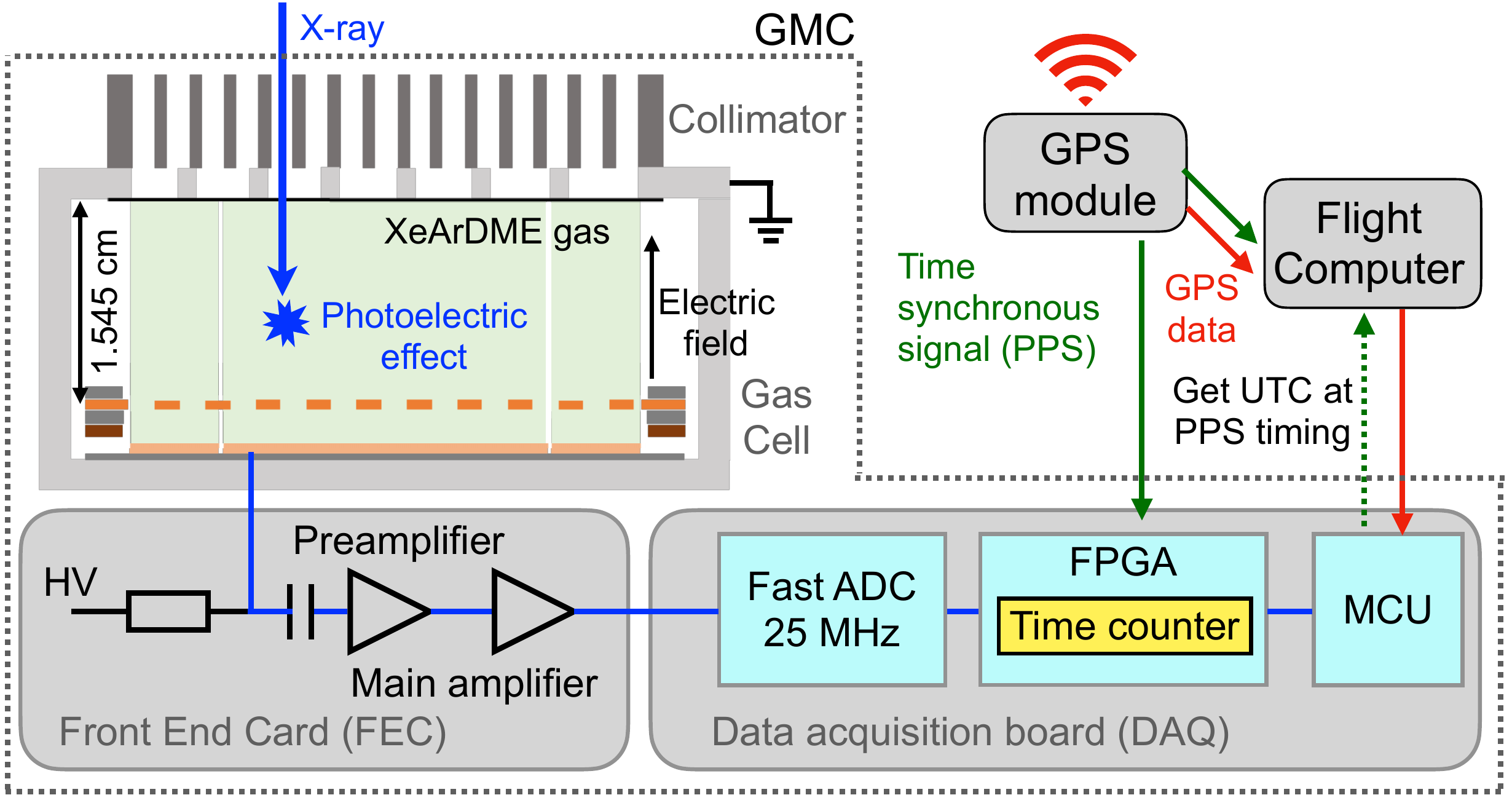}
\end{center}
\caption[example] 
{\engrev{A schematic} of the X-ray timing measurement system of the GMC and time synchronization between the GMC local clock and GPS time. 
Blue arrows represent X-ray event data, red arrows represent PPS signals, and green arrows represent GPS data.
\label{fig:TimingSystemDiagram} }
\end{figure}

The GMCs record X-ray photons on an event-by-event basis.
\engrev{The detectors comprises a collimator with a field of view (FoV) of 2.1$^\circ$ in full width half maxumum (FWHM), gas cell enclosing Xe, Ar, and dimethyl ether mixtured gas, front end card (FEC) for analog signal processing and application of high voltage to the gas cell, and data acquisition (DAQ) board for digital signal processing and communication with a PC and FC.}
The method of X-ray timing measurement is illustrated in Fig.~\ref{fig:TimingSystemDiagram}. 
An X-ray absorbed via the photoelectric effect in the gas cell produces a charge (electrons) signal, which is then collected at the electrodes.
\engrev{The charge signal is first converted into a voltage signal by a preamplifier on the FEC and then adjusted in amplitude by the main amplifier. 
Finally, its waveform is digitized by an analog-to-digital converter (ADC) operating at a sampling rate of 25~MHz on the DAQ board.}
After analog signal processing, the rise time of the X-ray signal is typically a few hundred nanoseconds.
\engrev{Timing measurement is performed using a real-time counter implemented on a field-programmable gate array (FPGA).}
The FPGA triggers the signal when the sampled ADC value is higher than the configured threshold.

The location of photoelectric absorption within the gas cell influences the absolute timing measurement.
The sensitive region of the GMC gas cell has a thickness of 1.545~cm.
The electric field applied to the sensitive regions of the GMC1 and GMC2 is 390 and 386~V/cm, respectively.
Under these conditions, the electron drift velocity is estimated to be approximately 1.4~cm/$\mu$s, corresponding to a maximum drift time of \engrev{approximately} 1.1~$\mu$s for photoelectric events.
Therefore, the contribution to the uncertainty of the absolute timing measurement is less than this value.

The FPGA appends a time counter value corresponding to the trigger timing to each event and transmits it, together with the digitized waveform data, \engrev{to a microcontroller unit (MCU) onboard the DAQ board via Serial Peripheral Interface.}
\rev{
The MCU stores the 40-bit time counter value with a time resolution of $\sim$61~$\mu$s, where the resolution is defined such that 14 bits correspond to 1~s.
To reduce the event data size, only the lower 16 bits of the 40-bit time counter are recorded for each detected X-ray event. 
Since the 16-bit event time counter wraps around exactly every 4~s, a carry-over flag is appended to the event data at each rollover.
To preserve the absolute timing information beyond this 4~s span, the upper 24 bits of the 40-bit time counter are recorded separately as ``carry data,'' which are updated at fixed 4~s intervals, corresponding to each increment of the upper 24-bit counter.
On the ground, the complete 40-bit time counter value for each event is reconstructed by combining the 16-bit event time with the 24-bit counter in the most recent carry data. 
When an event with a carry-over flag is encountered, the reference carry data are updated accordingly to ensure continuity of the reconstructed absolute event times.
}

\engrev{The GPS receiver generates and distributes pulse-per-second (PPS) signals to the FC and transmits position and the absolute time defined in Universal Time Coordinated (UTC) data at PPS timing to the FC.
As the GMC does not have its absolute time information, PPS signals are also distributed to the GMC from the GPS receiver to ensure synchronization.
For each PPS signal, the FPGA latches the time counter value with a resolution of 15~$\mu$s and transmits it to the MCU. 
Upon receiving the value, the MCU requests the UTC at the last PPS timing from the FC.
The MCU outputs GPS--PPS data comprising the time counter value and UTC at PPS timing.
Summarily, the GMC outputs three types of data: X-ray event, carry, and GPS--PPS data.}
DAQ transmits these data to the PC, and the data streams are downlinked to the ground.

On the ground, absolute times are assigned to X-ray events by linearly interpolating between two adjacent GPS-PPS data points for each event~\cite{ota2024automatic}. 
Let $C_{\rm evt}$ be the GMC time counter value assigned to an X-ray event, $T_1$ and $T_2$ be the UTC at PPS timing before and after the event, \engrev{respectively, and $C_1$ and $C_2$ be the GMC time counter values at PPS timing, the absolute time  $T_{\rm evt}$ of the X-ray event is calculated by linear interpolation as}
\begin{equation}
T_{\rm evt} = T_1 + \cfrac{T_2-T_1}{C_2-C_1}\times (C_{\rm evt}-C_1).
\end{equation}

\subsection{Data Reduction}
\label{subsec:DataReduction}
To produce cleaned event data, following the method described in Takeda et al. (2025), non-X-ray background signals and electrical noise were removed based on the characteristics of the waveforms. 
If the time counter value in the \engrev{event data} rolls over without being accompanied by a carry-over flag, the reconstruction of the time counter value cannot \engrev{be correctly performed}.
Such a case is identified as a bad time interval and excluded from the analysis.
\engrev{Good time intervals (GTIs) were defined as the periods during which high voltage was applied to the GMC, a target celestial object was in the FWHM of the FoV, GPS--PPS data were received at 1~s intervals, and GPS satellites were in view of the GPS receiver (nominally $\sim$10).}

\subsection{Verification of Timing Measurement}
\label{subsec:verification}
\begin{figure} [ht]
\begin{center}
\includegraphics[width=1\linewidth]{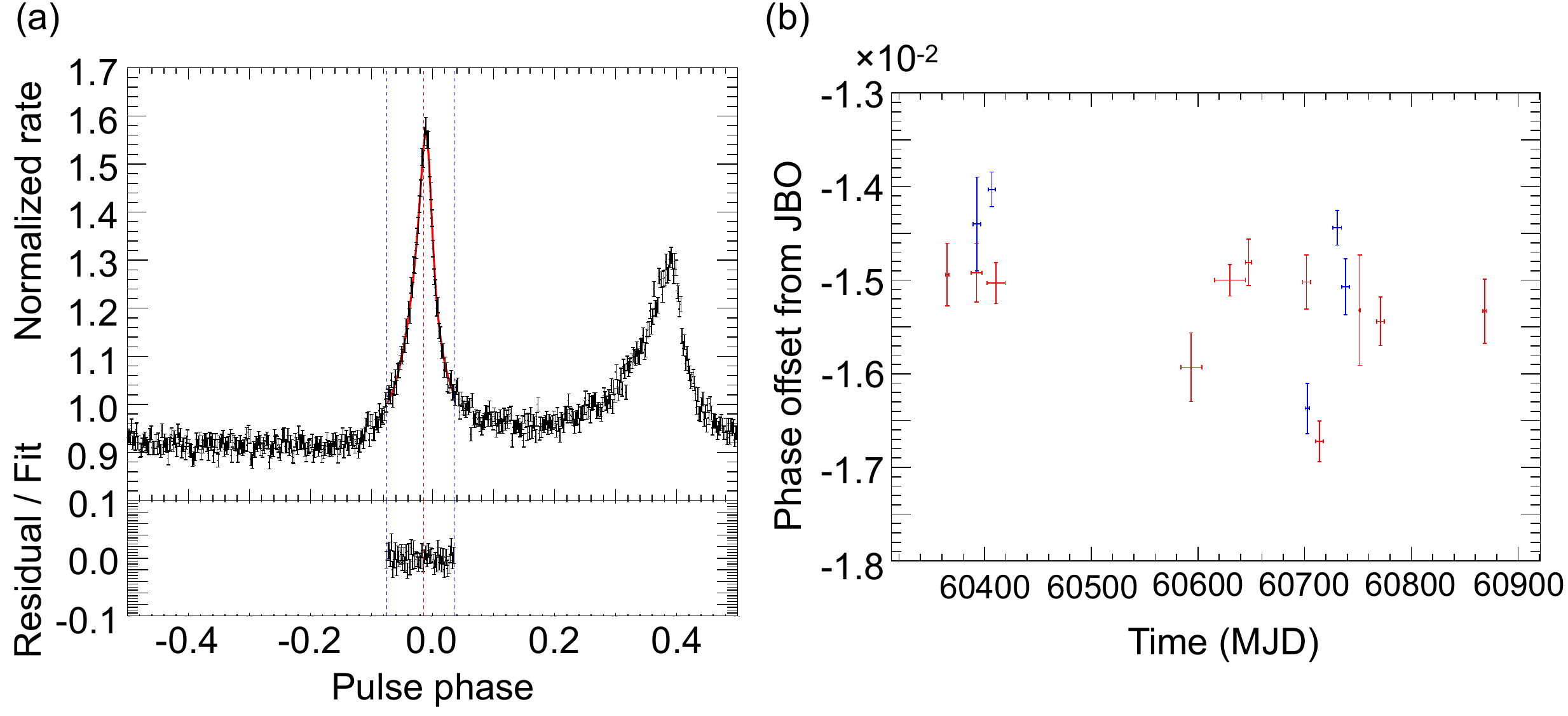}
\end{center}
\caption[example] 
{(a, top) \engrev{Pulse profile of the Crab Pulsar, created by epoch folding in the spin period.
Data were acquired by the GMC2 with an exposure time of 80~ks.}
The pulse phase was assigned to each X-ray photon based on the JBO radio observations.
\engrev{Blue dot lines represent the fitting range by Nelson's formula.
The red dot line is a fit result $\phi_0$.}
(a, bottom) Fit residuals normalized by the fitted values.
(b) Primary peak phase offset from JBO derived by the fit results of pulse profiles with Nelson's formula.
\engrev{Red and blue points represent the GMC1 and GMC2 data, respectively.}
Vertical error bars are 1-$\sigma$ error of the mean of the primary peak fit.
$1 {\rm phase} \simeq 33.83~{\rm ms}$.
Horizontal error bars are the start and end of each observation.
\label{fig:TimingVerification_radio} }
\end{figure}

\engrev{We verified the absolute timing capability of NinjaSat using the well-established radio timing ephemeris of the Crab Pulsar provided by the Jodrell Bank Observatory (JBO)~\cite{jbo}.}  
We evaluated timing accuracy by measuring the phase offset of the X-ray main peak relative to the JBO radio reference.
For the barycentric correction, we used the orbital files generated by the Attitude Determination and Control Subsystem (ADCS).  
The ADCS provides the Cartesian position and velocity data in the True Equator Mean Equinox (TEME) frame.  
For comparison with the JBO radio ephemeris and for subsequent barycentric corrections, these data were transformed from TEME to \engrev{the} J2000 coordinate system.
The photon phases of the Crab Pulsar X-ray events detected by the GMCs were then assigned using the \texttt{photonphase} command of the PINT software package~\cite{Luo_2021}, which performs both barycentric correction and pulsar phase calculation.  
The analysis was performed using the Crab Pulsar coordinates of (R.A., Dec.) = (83.633218, 22.014464) on the DE200 solar system ephemeris.
Using the Crab Pulsar spin frequency $\nu$, its first derivative $\dot{\nu}$, published by the JBO, and the second derivative $\ddot{\nu}$, calculated from the timing model, the pulsar phase $\phi(t)$ at a given barycentric time $t$ is computed as
\begin{equation}
\phi(t) = \nu\times(t-t_0) + \frac{1}{2}\dot{\nu}\times(t-t_0)^2 + \frac{1}{6}\ddot{\nu}\times(t-t_0)^3,
\end{equation}
where $t_0$ is defined at the time when $\phi(t_0)=0$.
Figure~\ref{fig:TimingVerification_radio}(a) shows an example of a folded pulse profile generated by assigning pulsar phase information to the X-ray event data with an exposure time of 80~ks observed by the GMC2 in February 2025.
We confirmed the primary peak phase around 0.0 and the secondary peak around 0.4, known as \engrev{the} Crab pulse shape.

To quantify the timing accuracy, we fitted the folded X-ray pulse profile with Nelson’s formula\cite{Nelson1970-dp}, \engrev{which comprises a sum of asymmetric harmonics, enabling a precise characterization of the pulse shape, particularly the sharp main peak.}
The X-ray counts at $\phi$, $L(\phi)$, is described as
\begin{equation}
L(\phi) = N\frac{1+\alpha_1(\phi - \phi_0)+\alpha_2(\phi - \phi_0)^2}{1+\alpha_3(\phi - \phi_0)+\alpha_4(\phi - \phi_0)^2}{\rm exp}\{-\beta(\phi - \phi_0)^2\}+l,
\label{eq:nelson}
\end{equation}
where $N$ is the peak height, $\alpha_1$, $\alpha_2$, $\alpha_3$, $\alpha_4$ and $\beta$ are \engrev{the} shape coefficients, $\phi_0$ is the peak phase, and $l$ is the off-pulse intensity level.
We fitted equation \ref{eq:nelson} to the primary peak of the folded pulse profile generated from all Crab Pulsar data observed between February 2024 and April 2025, after screening the data to include only the time intervals during which GPS data were continuously acquired at \engrev{1-s} intervals.
\engrev{Its fit range is $\phi=-0.075$ to $\phi=+0.0355$, and the statistical method is $\chi^2$-test.}
We fixed shape coefficient parameters to values used in the previous study with Proportional Counter Array onbard RXTE as $\alpha_1 = -30.79$, $\alpha_2 = 1550.04$, $\alpha_3 = -55.03$, $\alpha_4 = 4521.40$, and $\beta = 568.32$~\cite{Ge_2016}, and treated $N$, $\phi_0$, and $l$ as free parameters.
In the example shown in Fig.~\ref{fig:TimingVerification_radio}(a), the folded pulse profile was fitted using Nelson’s formula. The fit residuals were less than 10\%, indicating good agreement with the data. 
As a fit result of the example, the phase offset of the primary peak relative to the JBO ephemeris was determined as $\phi_0=(-1.44\pm0.02)\times10^{-2}$.
Figure~\ref{fig:TimingVerification_radio}(b) shows the pulse phase offsets from the JBO radio ephemeris derived from the fitting results for each NinjaSat observation data.
Throughout the observation period, the results from NinjaSat show that the pulse phases are consistently advanced from $1.4\times10^{-2}$ to $1.7\times10^{-2}$ relative to the JBO ephemeris. 
\engrev{
The average of the best-fit mean values of the phase offset obtained from fitting using Nelson's formula over the entire data set is \rev{$(-1.5 \pm 0.1)\times10^{-2}$}, where the quoted uncertainty is derived as the square root of the sum of the squared individual fit errors.
}
It corresponds to an offset of \rev{$(-5.1 \pm 0.4)\times10^2~\mu{\rm s}$}, where the conversion from phase to time with spin frequency of the monthly JBO ephemeris.
\engrev{
The standard deviation of the best-fit mean values is \rev{$0.1\times10^{-2}$} in phase, corresponding to \rev{$0.2\times10^2~\mu{\rm s}$} in time.
}
\rev{ 
This phase offset of $\sim 500~\mu$s between X-ray and radio is consistent with recent absolute timing studies by X-Ray Imaging and Spectroscopy Mission (XRISM) and NICER that directly compare X-ray pulse phases with the radio reference~\cite{sawada2025absolute}.
}

\begin{figure} [ht]
\begin{center}
\includegraphics[width=0.7\linewidth]{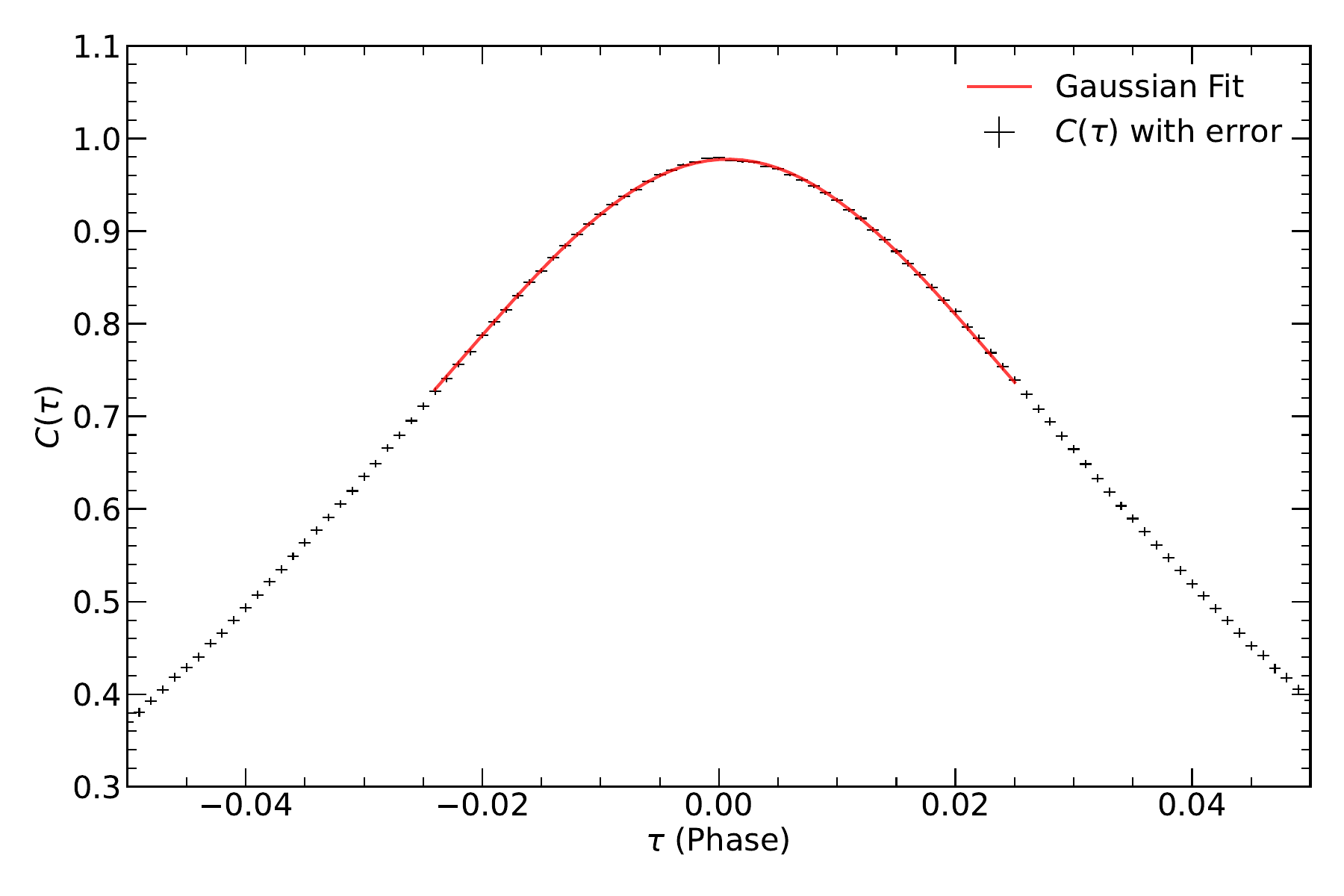}
\end{center}
\caption[example] 
{Example of cross-correlation of pulse profiles of NinjaSat and NICER as a function of the phase lag $\tau$. 
Black points are $C(\tau)$ calculated with data observed by NinjaSat and NICER in February 2025.
A red curve is a fitted Gaussian.
\label{fig:2025_Feb_cross_correlation_with_error} }
\end{figure}

To verify the absolute timing of NinjaSat more precisely, we compared its Crab Pulsar observations with the data obtained by NICER simultaneously in the 2--12~keV band.  
\engrev{For each observation data value, we generated pulse profiles using the JBO Crab Pulsar ephemeris and calculated the cross-correlation between the pulse profiles of NinjaSat and NICER.}
The phase of the folded profile is denoted by $\phi$, with the NinjaSat profile represented as $F(\phi)$, its average rate as $\bar{F}$, the NICER profile as $G(\phi)$, and its average rate as $\bar{G}$.  
The cross-correlation function was then evaluated by shifting the NinjaSat profile by a phase lag $\tau$ as
\begin{equation}
C(\tau) = \cfrac{\sum_{j} \left[F(\phi_j-\tau) - \bar{F}\right] \left[G(\phi_j) - \bar{G}\right]}{\sqrt{\sum_{j}\left[F(\phi_j-\tau) - \bar{F}\right]^2}\sqrt{\sum_{j}\left[G(\phi_j) - \bar{G}\right]^2}}.
\end{equation}
The function $C(\tau)$ was calculated\engrev{,} and its maximum was determined by fitting the peak region with a Gaussian function, as shown in Fig.~\ref{fig:2025_Feb_cross_correlation_with_error}.  
\engrev{The best-fit mean values obtained from the Gaussian fitting} provided an estimate of the relative time offset between NinjaSat and NICER.
We performed the analysis for six observation epochs between 2024 and 2025. 
The measured offsets between NinjaSat and NICER ranged from approximately $–15$ to $38~\mu$s. 
\engrev{
These results are presented in Table~\ref{tab:timing_calibration}.
The average of the best-fit mean values with the square root of the sum of the squared individual fit errors of $C(\tau)$ is \rev{$8\pm1~\mu$s}, and the standard deviation of the best-fit mean values is \rev{$21~\mu$s}.
}

\begin{table}[t]
\centering
\begin{threeparttable}[t]
\centering
\caption{Timing offsets of NinjaSat relative to NICER using Crab Pulsar observations.
}
\label{tab:timing_calibration}
\begin{tabular}{p{0.8cm}p{2cm}p{2cm}p{2.5cm}p{2cm}p{2cm}r}
\hline
GMC\tnote{a} &NinjaSat start (MJD)\tnote{b} &NinjaSat Exp. (ks)\tnote{c}& NICER OBS~ID & NICER start (MJD)\tnote{d} &NICER Exp. (ks)\tnote{e}& Offset ($\mu$s)\tnote{f} \\
\hline
1 &60363.4 & 27.67& 6013010130 &60364.1&1.37& $-14.1 \pm 1.0$ \\
1 &60387.5 & 23.35& 7013010101 &60388.2&1.16 & $11.5 \pm 2.0$\\
1 &60698.0 & 39.02& 7013010128--7013010133, 7013010135, 7013010136 &60698.8&2.16& $-15.1 \pm 1.0$ \\
2 &60700.5 & 41.47& 7013010128--7013010133, 7013010135, 7013010136 & 60698.8&2.16& $38.2 \pm 1.1$\\
1 &60710.4 & 56.47& 7013010137, 7013010138, 7013010139, 7013010140& 60708.5&0.81& $3.2 \pm 1.6$\\
% 1 &60710.4 & 56.47&  7013010138, 7013010139, 7013010140& 60708.5&0.81& $3.2 \pm 1.6$\\
2 &60726.4 & 78.85& 7013010141, 7013010142& 60727.2&2.63& \rev{$26.0 \pm 1.0$}\\
\hline
 & & & & & Average& $8.3 \pm 1.3$\\
 & & & & & SD\tnote{g}& \rev{21}\\
\hline
\end{tabular}
\begin{tablenotes}
\footnotesize
\item[a]{Used GMC for simultaneous observation.}
\item[b]{
Start date of Crab Pulsar observation performed by NinjaSat, \engrev{expressed} in Modified Julian Date (MJD).
The observation periods of NinjaSat and NICER overlapped at least in part, with any non-overlapping intervals being shorter than one week.}
\item[c]{
Exposure time of observation performed by NinjaSat.}
\item[d]{
Start date of Crab Pulsar observation performed by NICER, \engrev{expressed} in MJD.}
\item[e]{
Exposure time of observation performed by NICER.}
\item[f]{
The pulse offset of NinjaSat relative to NICER obtained from the cross-correlation analysis, with error values representing the uncertainties of the fitted mean value.}
\item[g]{\engrev{SD is the standard deviation of the best-fit mean values of pulse offsets.}}
\end{tablenotes}
\end{threeparttable}
\end{table}

\section{Pulsar Navigation Method}
\label{sec:xnav_method}

\subsection{The Orbit Dynamics}
\label{subsec:OrbitalDynamics}
\begin{figure} [ht]
\begin{center}
\includegraphics[width=0.5\linewidth]{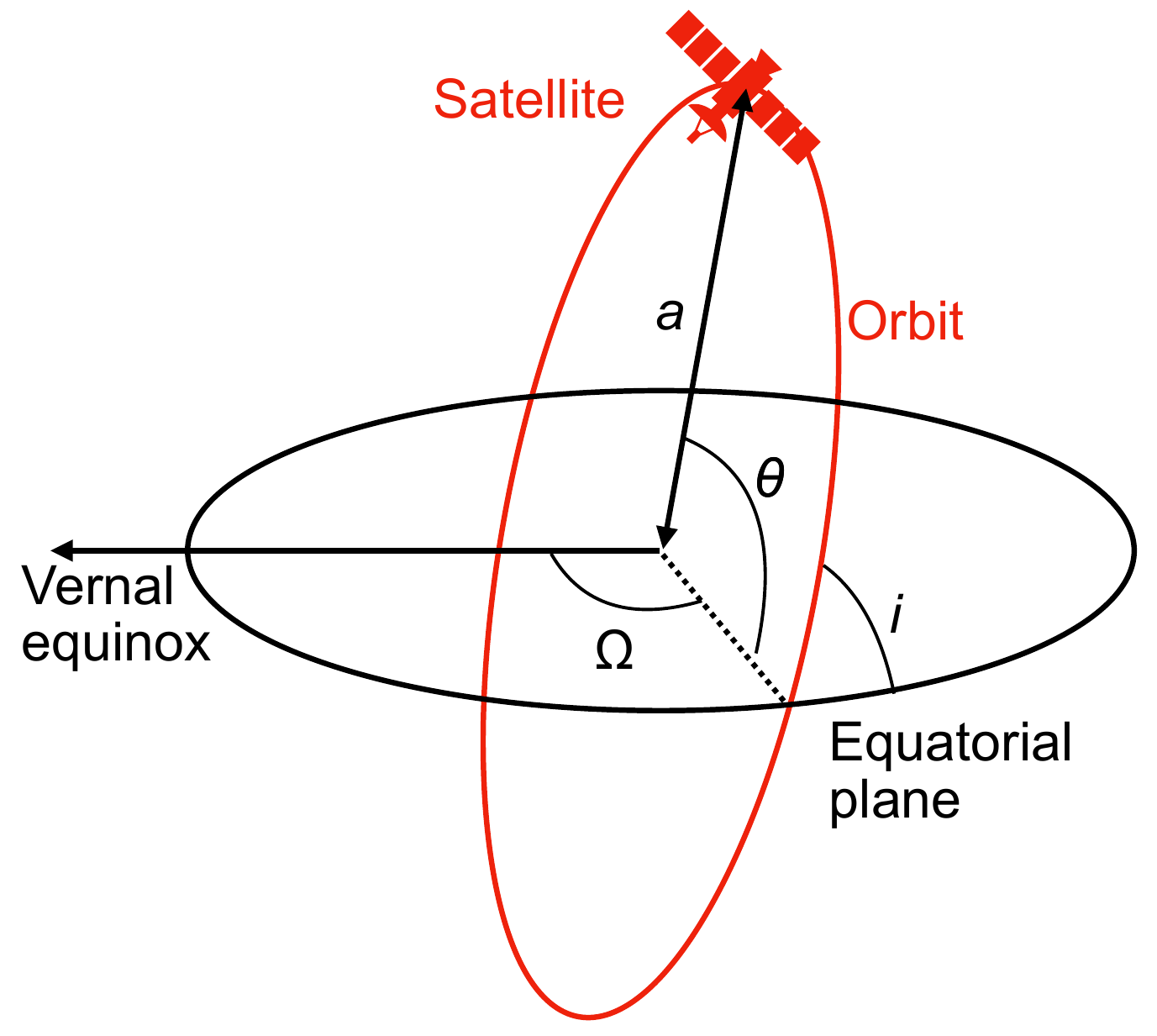}
\end{center}
\caption[example] 
{Orbital elements for a circular orbit around the Earth.
\engrev{$a$ is the orbital radius, $i$ is the inclination, $\theta$ is the orbital phase at the start of the observation, and $\Omega$ is the right ascension of the ascending node.}
\label{fig:OrbitDynamics} }
\end{figure}

As an orbital propagation algorithm, we used the Simplified General Perturbations 4 (SGP4) model~\cite{vallado2006revisiting}, 
which is the standard analytical orbit propagator used with Two-Line Element sets (TLEs) to predict satellite positions and velocities. 
We assumed that NinjaSat follows a circular orbit, as its eccentricity is almost \engrev{less than} 0.001, which is negligible for the accuracy required in this study. 
\engrev{As shown in Fig.~\ref{fig:OrbitDynamics}, a circular orbit can be described by four parameters: the orbital radius $a$, inclination $i$, ascending node $\Omega$, and orbital phase at the start of the observation $\theta$.} 
The orbital radius $a$ is related to the mean motion $n$ through the gravitational constant of Earth $\mu = 3.9860\times10^{14}$~m$^3$/s$^2$$=a^3n^2$~\cite{nima:2000}. 
\engrev{Furthermore, atmospheric drag was also considered, and the atmospheric drag term $B$ was included as an orbital parameter of NinjaSat.}
In this configuration, the eccentricity $e$ and the argument of perigee $\omega$ were fixed to zero.
In this study, the orbital calculations were performed by inputting $B$, $i$, $n$, $\theta$, and $\Omega$ into the SGP4 model. 

\subsection{The Optimization of Orbital Elements}
\label{subsec:OptMethod}
\engrev{In the demonstration performed with Insight-HXMT, orbital elements were optimized by varying each parameter individually~\cite{Zheng_2019}. 
However, the spacecraft position arises from the interplay of several orbital elements, and therefore, constraining some of them to fixed values may limit the validity of the SEPO method.
Simultaneous optimization of multiple orbital elements is desirable.
We formulated the estimation of orbital elements from pulsar observations as a Bayesian optimization problem in which the objective function is defined as the sharpness of the pulse profile obtained via epoch-folding analysis of the Crab Pulsar observation data. 
We considered $B$, $i$, $n$, $\theta$, and $\Omega$, which are used as inputs to the SGP4 propagator, as the optimization parameters.  
In this framework, each iteration of the optimization loop involved the following procedures:}  

Firstly, a trial set of orbital elements was selected. 
The initial orbital elements were derived from \engrev{TLEs} released by the North American Aerospace Defense Command (NORAD) archived on CelesTrak~\cite{celestrak}, which were closest in time to the start of the pulsar observation.
\engrev{Ten random sample sets were included in the initial values to efficiently explore the parameter space.}
Subsequent candidates were proposed by the Bayesian optimization algorithm. 
Using these elements, the satellite position was \engrev{determined} through propagation with the SGP4 model at an interval of 10~s.

\engrev{Second, the barycentric correction of observed event data was performed using satellite positions calculated from the candidate orbital elements using the SGP4 model. 
Then, the corrected photon arrival times were assigned pulsar phases with the JBO Crab Pulsar ephemeris and folded to produce a pulse profile.}

\engrev{Finally, the sharpness of a pulse profile was quantified. 
$P(\phi_i)$ represents the count in a phase bin $\phi_i$, and $\bar{P}$ represents their mean value. 
The pulse significance was quantified by}
\begin{equation}
\chi^2 = \sum_j \frac{(P(\phi_j)-\bar{P})^2}{\bar{P}}.
\label{eq:chi^2}
\end{equation}
The value $- \chi^2$ was returned to the Bayesian optimization routine as the evaluation of the current parameter set. 

Bayesian optimization was performed using the GPyOpt library~\cite{gpyopt2016}. 
\engrev{An acquisition function of the Bayesian optimization is the lower confidence bound, expressed as $\mu-2\sigma_{\rm g}$, where $\mu$ and $\sigma_{\rm g}$ are the predicted mean and standard deviation estimated by the Gaussian process model, respectively.
We used the Matern52 for the Gaussian process regression model~\cite{williams2006gaussian}.  
The kernel function between the two parameter vectors $x$ and $x'$ is defined as}
\begin{equation}
k(x,x') = \sigma_k^2 \left(1 + \sqrt{5}r + \frac{5}{3}r^2\right) \exp\left(-\sqrt{5}r\right),
\end{equation}
where 
\begin{equation}
r = \sqrt{\sum_{d=1}^{5} \frac{(x_d - x'_d)^2}{\ell_d^2}}.
\end{equation}
The value $\sigma_k^2$ is the output variance and $\ell_d$ is the characteristic length scale for the $d$-th dimension.
We configured the initial values of $\ell_d$ to be the search ranges of each parameter.

\section{Observational and Reference Data}
\label{sec:ObsAndData}

\begin{figure} [ht]
\begin{center}
\includegraphics[width=0.9\linewidth]{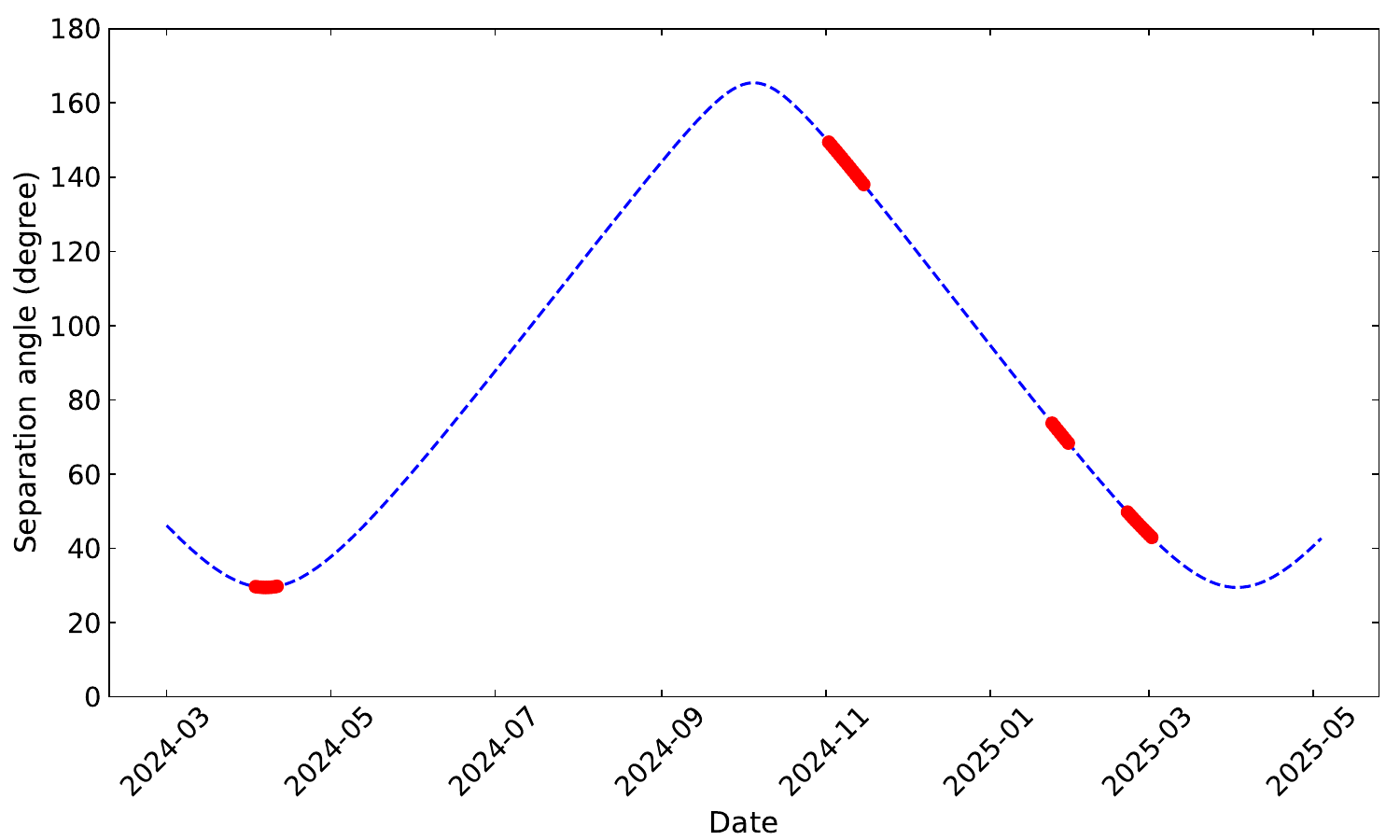}
\end{center}
\caption[example] 
{\rev{Time-series variation in the separation angle between the normal vector of the NinjaSat orbital plane and Crab Pulsar direction.
The blue dashed line represents the separation angle over the entire NinjaSat operational period.
The red solid line indicates the time intervals when NinjaSat observed the Crab Pulsar.
}
\label{fig:OrbPlaneAngYear} }
\end{figure}

In the SEPO method, satellite positioning is performed based on the pulse significance of a single celestial source. 
\engrev{As NinjaSat is in an SSO, the angle between the orbital plane and target source changes with the Earth's revolution. This may affect positioning accuracy depending on the observation epoch.} 
\engrev{Figure~\ref{fig:OrbPlaneAngYear} shows a calculation result of the time-series variation in the angle between the normal vector of the NinjaSat orbital plane and Crab Pulsar direction.}
To evaluate this effect, we \engrev{performed} four separate observations at different epochs.
To investigate the dependence of navigation accuracy on the observation period, we conducted long-term observations of the Crab Pulsar in April 2024, November 2024, January 2025, and February 2025. 
\engrev{These epochs were chosen to represent different geometrical configurations between the orbital plane of NinjaSat and Crab Pulsar direction, as the angle varies seasonally with the revolution of Earth around the Sun. 
The details of the exposure time, number of events detected, and angular separation between the orbital plane and Crab are presented in Table~\ref{tab:season_obs}.}

\begin{table}[t]
\centering
\begin{threeparttable}[t]
\centering
\caption{Summary of Crab Pulsar observations used for demonstration of in-orbit pulsar navigation.}
\label{tab:season_obs}
\begin{tabular}{lcccc}
\hline
Start time& Obs. span (ks)\tnote{a} & Exp. (ks)\tnote{b} & Number of events ($\times10^6$) & Ang. Dist. ($^{\circ}$)\tnote{c} \\
\hline
2024-04-02\\11:25:59 & 780 & 130 & 3.24 & 29--30 \\
2024-11-01\\12:00:20 & 1190 & 97  & 2.25 & 138--149 \\
2025-01-23\\01:09:53 & 645 & 80  & 2.13 & 68--74 \\
2025-02-20\\10:47:33 & 874 & 104 & 3.14 & 43--50 \\
\hline
\end{tabular}
\begin{tablenotes}
\footnotesize
\item[a]{Obs. span is the time from the first GTI start to the last GTI end (including gaps).
The angular separation is defined between the orbital \rev{normal plane} and the Crab Pulsar direction.}
\item[b]{Exp. is \rev{the} exposure time of observation.}
\item[c]{Ang. Dist. is the angular distance between \rev{the Crab Pulsar direction and normal vector of the orbital plane of NinjaSat.}}
\end{tablenotes}
\end{threeparttable}
\end{table}

\engrev{The JBO Crab Pulsar ephemeris used for epoch folding at each observation epoch are presented in Table~\ref{tab:jbo_ephem}.}
The initial orbital elements derived from the corresponding NORAD TLEs, which served as the starting values for the Bayesian optimization, are \engrev{presented} in Table~\ref{tab:tle_init}.
In the NORAD TLE format, the orbital elements are expressed for an elliptical orbit, and both the $\omega$ and \engrev{mean anomaly} ($M_0$) at the epoch are specified.  
In this study, \engrev{as} a circular orbit was assumed, the orbital phase at the epoch time was defined as $\theta$, measured from the equatorial plane, and calculated as the sum of the $\omega$ and $M_0$ in \engrev{TLEs}.

We determined the search ranges of \engrev{the} orbital elements for \engrev{the} Bayesian optimization, referring to previous studies with large satellites, while expanding them to account for the greater orbital variability expected for a small satellite such as NinjaSat.  
We defined ranges an order of magnitude broader than those used in previous \engrev{studies}.  
For $i$ and  $\Omega$, Insight-HXMT and POLAR set the search ranges of $\pm0.5^\circ$ and $\pm1.0^\circ$, respectively.
In our analysis, we defined the search range of $i$ as $\pm5^\circ$ and $\Omega$ as $\pm10^\circ$.  
\engrev{For the orbital phase $\theta$, corresponding to $M_0$, the search range of the previous study was $\pm1.0^\circ$, and we defined the search range of $\theta$ as $\pm10^\circ$.  
Previous studies considered the semi-major axis $a$ as an orbital parameter.
The search ranges of studies on Insight-HXMT and POLAR are$\pm200$~m and $\pm50$~m, respectively.  
As $n=\sqrt{\mu a^{-3}}$, the perturbation can be expressed as $\Delta n = -3n\Delta a/2a$.}  
With $n \simeq 0.066$~rad/min for NinjaSat, a semi-major axis variation of $\Delta a = 1$~km corresponds to $\Delta n \sim 0.5\times10^{-5}$~rad/min.  
\engrev{Based on this scaling, the search range of $n$ is $\pm2\times10^{-5}$~rad/min.  
As $B$ was not considered a parameter in previous studies, we referred to values reported in the NORAD TLEs, which lie in the range $(0\text{--}2)\times10^{-3}~R_{\oplus}$$^{-1}$.  
We defined the search range $B$ as $\pm1\times10^{-3}$~$R_{\oplus}$$^{-1}$.
The parameter search ranges are presented in Table~\ref{tab:param_range}.}

\begin{table}[t]
\centering
\begin{threeparttable}[t]
\centering
\caption{JBO Crab Pulsar ephemeris parameters used for epoch folding at each observation epoch.}
\label{tab:jbo_ephem}
\begin{tabular}{lcccc}
\hline
Epoch & $\nu$ (Hz) & $\dot{\nu}$ ($\times 10^{-10}$~Hz/s) & $\ddot{\nu}$ ($\times 10^{-21}$~Hz/s$^2$) & $t_0$ (MJD) \\
\hline
2024 Apr & 29.5620206246 & -3.6666827& 9.0958342755 & 60415.0000002203\\
2024 Nov & 29.5552429346& -3.6647131& 9.0881486815 & 60629.0000001102\\
2025 Jan & 29.5533117036& -3.6640430& 9.0854190830 & 60690.0000001045\\
2025 Feb & 29.5523303683& -3.6637295& 9.0841660755 & 60721.0000001673\\
\hline
\end{tabular}
\begin{tablenotes}
\footnotesize
\item[]{}
\end{tablenotes}
\end{threeparttable}
\end{table}

\begin{table}[t]
\centering
\begin{threeparttable}[t]
\centering
\caption{Initial orbital elements defined with NORAD TLEs for Bayesian optimization.}
\label{tab:tle_init}
\begin{tabular}{lrrrrr}
\hline
Epoch & $B$ ($\times10^{-3}$~$R_{\oplus}^{-1}$)\tnote{a} & $i$ ($^{\circ}$)\tnote{b} & $\theta$ ($^{\circ}$)\tnote{c} & $n-0.066$ ($\times10^{-5}$~rad/min)\tnote{d} & $\Omega$ ($^{\circ}$)\tnote{e} \\
\hline
2024 Apr & \rev{0.68} & \rev{97.46} & 360.24 & \rev{ 29.90} & 169.45 \\
2024 Nov & \rev{2.04} & \rev{97.44} & 360.22 & \rev{ 88.06} & 21.42 \\
2025 Jan & \rev{1.46} & \rev{97.43} & 360.18 & \rev{125.04} & 104.34 \\
2025 Feb & \rev{1.23} & \rev{97.43} & 360.17 & \rev{139.30} & 134.03 \\
\hline
\end{tabular}
\begin{tablenotes}
\footnotesize
\item[a]{
$B$ is the atmospheric drag term.
Its unit $R_{\oplus}$ is Earth radius.
$R_{\oplus} = 6378.137~{\rm km}$ reffering WGS84 as a geodetic system~\cite{sgp4python}}
\item[b]{$i$ is the inclination.}
\item[c]{$\theta$ is the orbital phase at the start of the observation.}
\item[d]{$n$ is the mean motion.}
\item[e]{$\Omega$ is the ascending node.}
\end{tablenotes}
\end{threeparttable}
\end{table}

\begin{table}[t]
\centering
\begin{threeparttable}[t]
\centering
\caption{Parameter search ranges used in the Bayesian optimization of orbital elements, defined relative to the initial values.}
\label{tab:param_range}
\begin{tabular}{lc}
\hline
Parameter & Search range \\
\hline
$B$& $\pm 1.0 \times 10^{-3}$ $R_{\oplus}$$^{-1}$ \\
$i$& $\pm 5.0^{\circ}$ \\
$\theta$& $\pm 10^{\circ}$ \\
$n$& $\pm 2.0 \times 10^{-5}$ rad/min \\
$\Omega$& $\pm 10^{\circ}$ \\
\hline
\end{tabular}
\begin{tablenotes}
\footnotesize
\item[]{}
\end{tablenotes}
\end{threeparttable}
\end{table}

\section{Results}
\label{sec:xnav_results}
\begin{figure}[t]
\centering
\includegraphics[width=0.8\linewidth]{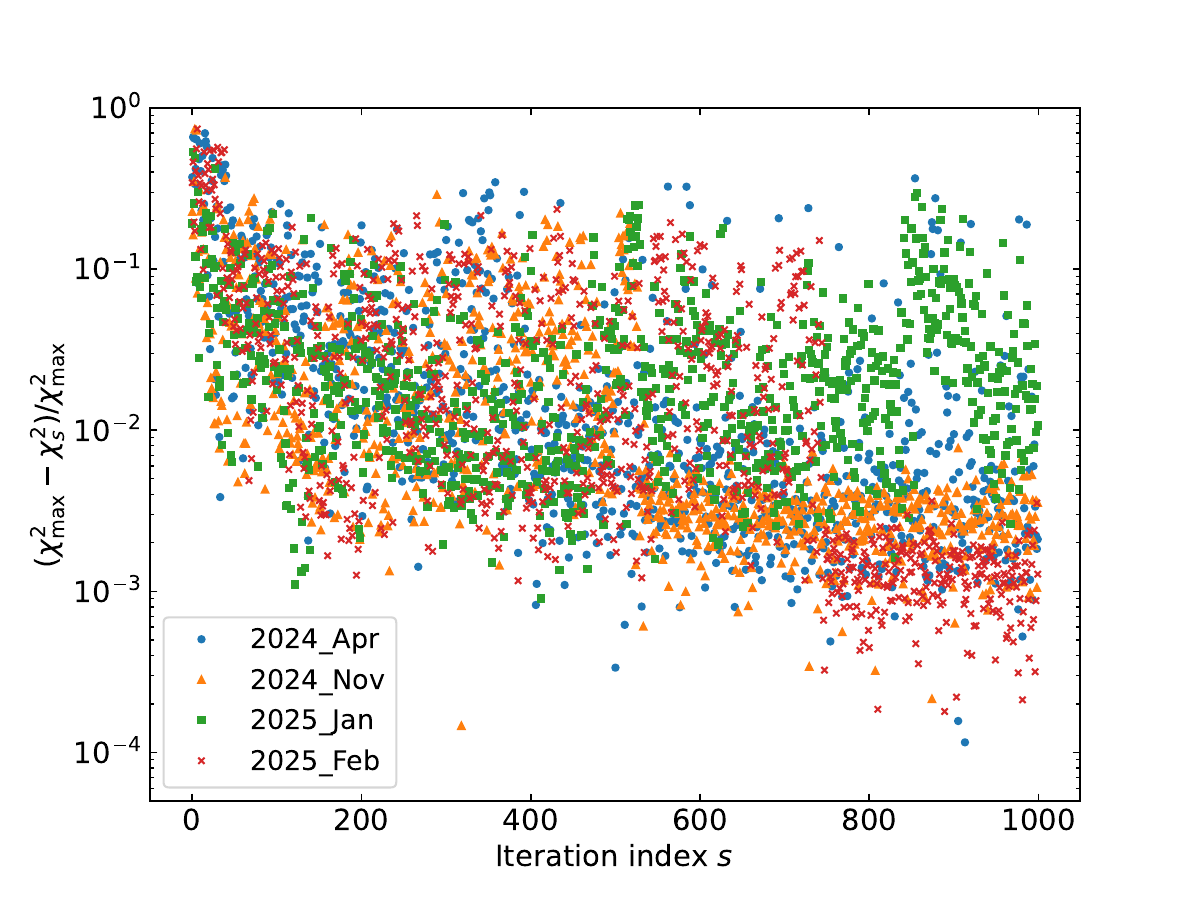}
\caption{
Instantaneous regret curves of \engrev{the} Bayesian optimization normalized by $\chi_{\rm max}^2$.
Blue, orange, green, and red points, respectively, indicate April 2024, November 2024, January 2025, and February 2025.
}
\label{fig:RegretPlot}
\end{figure}

To assess the degree of convergence and stability of the optimization process of \engrev{the} orbital elements, we visualized the evaluated objective function $\chi^2$ (Eq. (\ref{eq:chi^2})) and sampling points \engrev{using} the optimization process. 
\engrev{Figure~\ref{fig:RegretPlot} shows the residual between the evaluated objective function at an iteration number of $s$, $\chi_{s}^2$, and the maximum value of $\chi^2$, $\chi_{\rm max}^2$, normalized by $\chi_{\rm max}^2$ as a function of $s$.
This is generally referred to as an instantaneous regret curve~\cite{6138914}.}
In Bayesian optimization with Gaussian processes, the computational cost scales as $s^3$~\cite{williams2006gaussian}.
\engrev{Practically, the total computation time, including objective function evaluations, often increased nonlinearly if $s>1000$.}  
\engrev{Therefore, in this study, we set the termination criterion of the optimization as $s =1000$.}
By the end of 1000 iterations, the November 2024 and the February 2025 datasets had converged most stably, with variations \rev{less than 0.005, which corresponds to a variation of $\chi^2$ from the maximum value of less than 0.5\%.}
\engrev{By} contrast, the April 2024 and January 2025 datasets continued to sample values with variations exceeding 0.1 throughout the optimization, indicating unstable behavior.
\engrev{Orbital elements that are responsible for the fluctuations in the regret plot and the extent to which these elements affect the positioning accuracy have not been elucidated.}
Therefore, we investigated the characteristics of each orbital element and evaluated the positioning accuracy of the parameter sets of the $\chi_{\rm max}^2$.

\begin{figure} [ht]
\begin{center}
\includegraphics[width=1\linewidth]{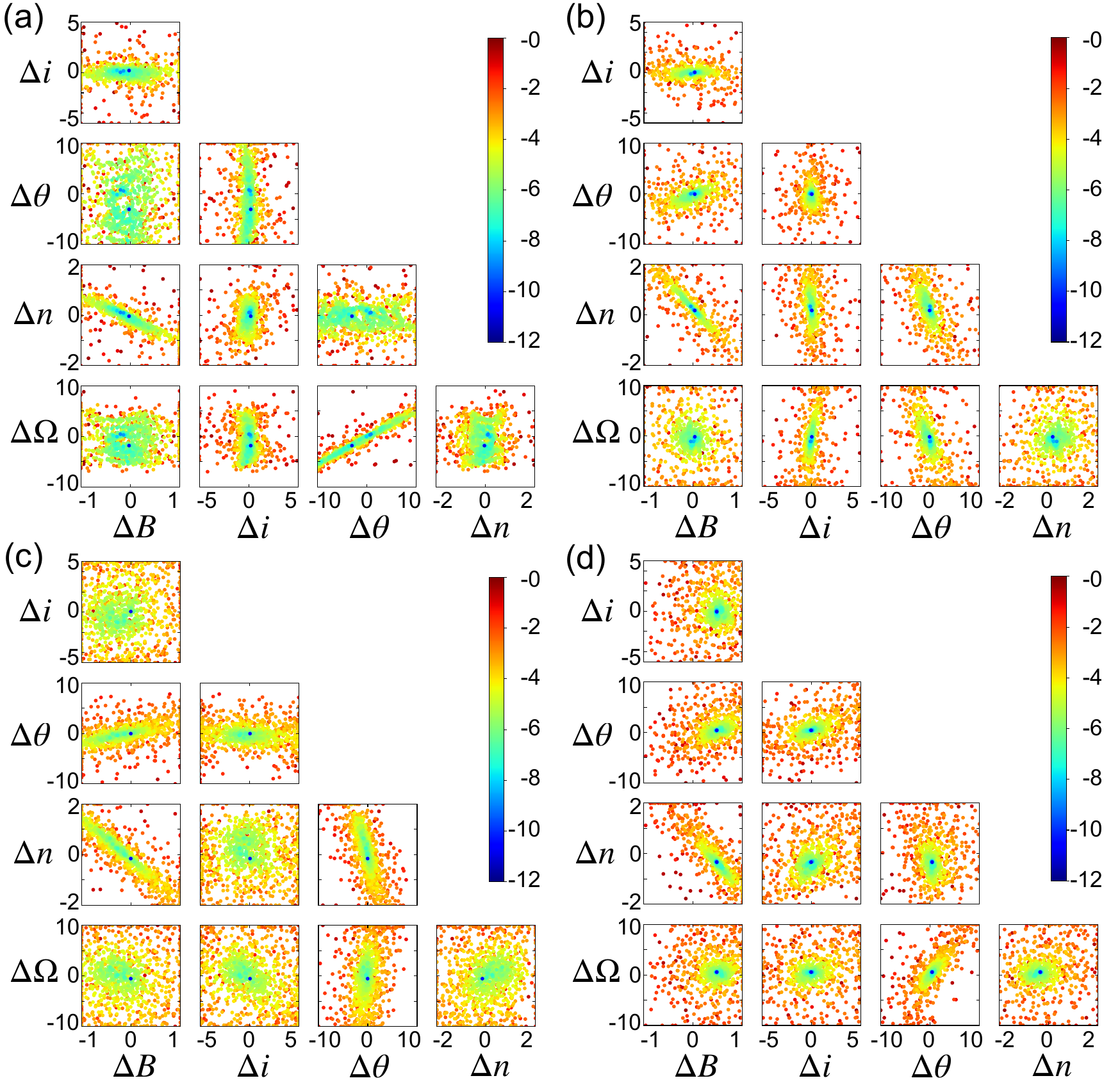}%\end{tabular}
\end{center}
\caption[example] 
{Distribution of sampled orbital elements from the Bayesian optimization for each observation epoch: (a) April 2024, (b) November 2024, (c) January 2025, and (d) February 2025.
The scatter plots show pairwise correlations between orbital parameters, where the $x$- and $y$-axes represent deviations from the initial values.
The color scale indicates the significance of the pulsar signal, \engrev{represented as $\ln\{(1+\chi^2_{\max}-\chi_s^2)/\chi^2_{\max}\}$.
The unit} of $B$ is $10^{-3}R_{\oplus}^{-1}$.
\engrev{The unit} of $i$, $\theta$, and $\Omega$ are degree ($^{\circ}$).
\engrev{The unit} of $n$ is $10^{-5}$rad/min.
\label{fig:BayesianSample} }
\end{figure}

Figure~\ref{fig:BayesianSample} shows the distributions of sampled points obtained in the Bayesian optimization analysis for the four observation datasets, illustrating the correlations between pairs of orbital elements.  
The value plotted on the $z$-axis is $\ln\{(1+\chi^2_{\max}-\chi_s^2)/\chi^2_{\max}\}$.
In the April 2024 dataset, the sampling exhibited a correlation between the orbital phase at the start of the observation $\theta$ and the ascending node $\Omega$.
\engrev{This tendency renders it challenging to determine a unique optimal solution.
By contrast, the inclination $i$ exhibits a narrow distribution, indicating that this parameter was more tightly constrained.  
The distributions of $i$ and $\Omega$ became broader in the January 2025 dataset than in the April 2024 dataset, rendering the significance less sensitive to changes in inclination.
However, the distribution of $\theta$ became narrower than in the April 2024 datasets.}
In the November 2024 and February 2025 datasets, the sampled points for all parameters were concentrated around a particular region, \engrev{indicating convergence of the optimization near a single solution.}
\engrev{Across all epochs, the sampled points of $B$ and $n$ were correlated.}

The best orbital elements obtained from \engrev{the} Bayesian optimization for the four observation epochs are \engrev{presented} in Table~\ref{tab:best_orbit}.
\rev{
As a reference measure for comparing the relative precision of the obtained orbital elements at each epoch, we list the deviations of the sampled values around the high-significance peak region shown in Fig.~\ref{fig:BayesianSample}.
We quantified these deviations as the standard deviation of the sampled values satisfying $(\chi^2_{\rm max}-\chi^2_{\rm s})/\chi^2_{\rm max} < 0.005$, which corresponds to samples with a $\chi^2$ variation of less than 0.5\% from the optimal sample.
The number of sampled points satisfying this criterion is 353, 506, 131, and 402 for the 2024 April, 2024 November, 2025 January, and 2025 February observations, respectively.
As also shown in Fig.~\ref{fig:BayesianSample}, the deviations of $\theta$ and $\Omega$ in April 2024 are notably larger than those in the other epochs.
}

\begin{figure} [ht]
\begin{center}
\includegraphics[width=1\linewidth]{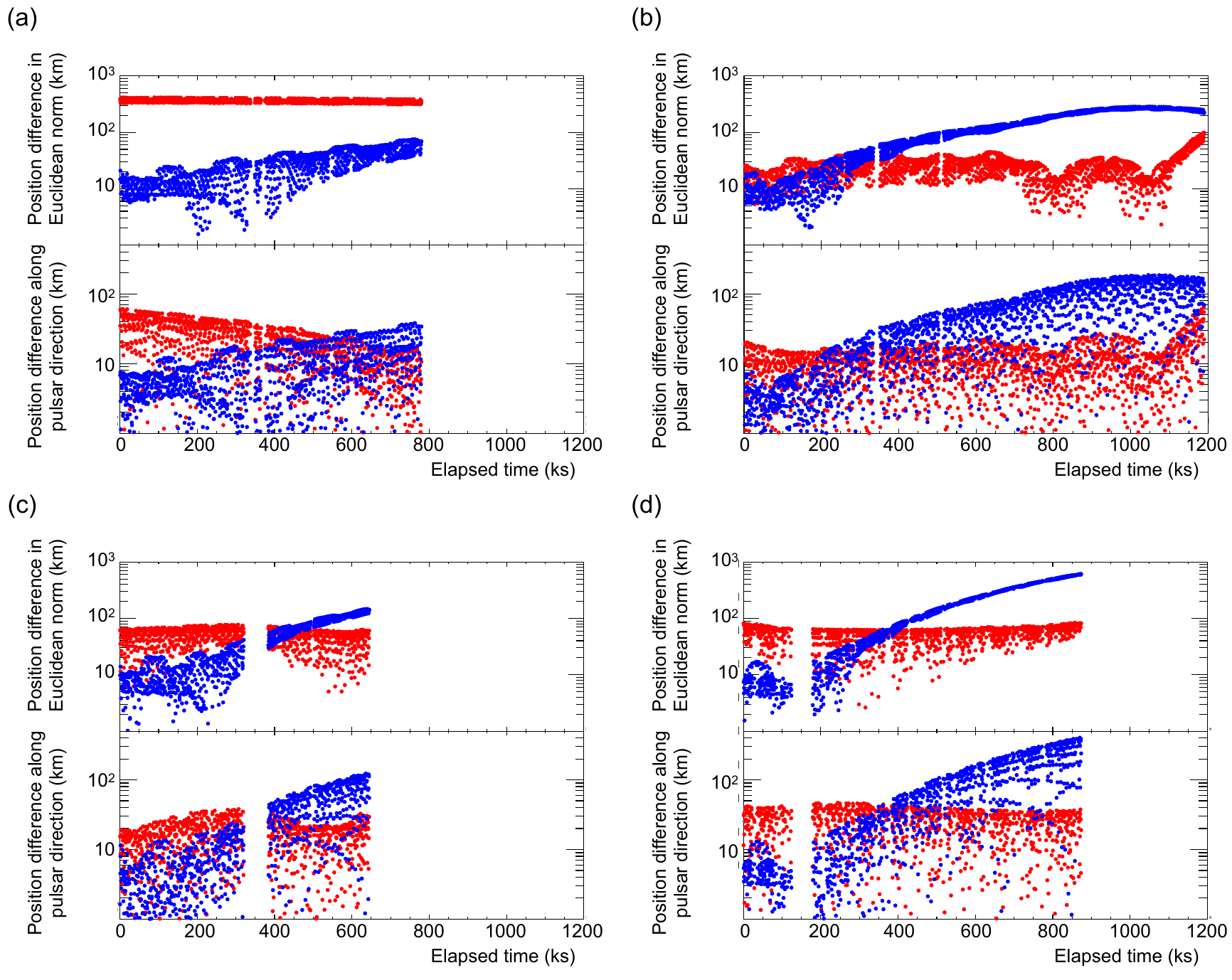}
\end{center}
\caption[example] 
{
\engrev{Position differences between the estimated and GPS positions: (upper panel) the Euclidean norm and (lower panel) the component along the Crab Pulsar direction, for (a) April 2024, (b) November 2024, (c) January 2025, and (d) February 2025.}
Red points show the position differences estimated using the SEPO method, while blue points show those estimated using the initial NORAD TLE. 
The horizontal axis represents the elapsed time since the start of the observation.
\label{fig:ResultPosDiff} }
\end{figure}

We evaluated the accuracy of the estimated orbits by comparing them with GPS-derived positions in the Earth-centered, Earth-fixed (ECEF) coordinate system.
\rev{Red points in upper panels of} Fig.~\ref{fig:ResultPosDiff}(a)--(d) show \rev{the three-dimensional (Euclidean norm)} position differences between the SEPO-derived and GPS-derived orbits for the four Crab Pulsar observations.
\rev{
Red points in lower panels of Fig.~\ref{fig:ResultPosDiff}(a)--(d) show the position differences projected onto the pulsar direction.
}
\rev{
For reference, we also plot the position differences between the GPS-derived positions and those derived from the initial NORAD TLE (Table~\ref{tab:tle_init}), shown as blue points in both the upper and lower panels of Fig.~\ref{fig:ResultPosDiff}(a)--(d).
The increase in the position difference of the blue points with elapsed time in Fig.~\ref{fig:ResultPosDiff} reflects the limited temporal validity of the initial NORAD TLE data.
The accuracy of the NORAD TLE–derived orbit degrades with time, and the resulting position error typically grows to several tens of kilometers within approximately one week ($\sim600$~ks).
On the other hand, the accuracy of the SEPO-derived orbit remains stable over time.
The spread of the data points arises from the orbital motion around the Earth.
The broader spread of the blue points in the lower panels compared to the upper panels is caused by periodic variations introduced when the position differences are projected onto the pulsar direction.
The spread of the red points in Fig.~\ref{fig:ResultPosDiff}(b)--(d) indicates that the position differences between the SEPO-derived and GPS-derived orbits exhibit oscillatory behavior, which is associated with periodic position variations caused by deviations in orbital plane parameters such as $i$ and $\Omega$.
}

\rev{
Table~\ref{tab:sepo_accuracy} presents the root mean square (RMS) time-series position differences between the GPS-derived positions and the orbit obtained using the optimal orbital elements derived from the SEPO method, evaluated over each observation period.
In addition, the table lists the standard deviations of the RMS position differences, evaluated from orbits generated using orbital element sets satisfying $(\chi^2_{\rm max}-\chi^2_{\rm s})/\chi^2_{\rm max} < 0.005$, to assess how the dispersion of the orbital elements propagates into position space.
}
\rev{
In April 2024, the Euclidean norm position difference reached $3.7\times10^2$~km with a standard deviation of $3.1\times10^2$~km, whereas the projected position difference along the Crab Pulsar direction was less than 40~km.
}
\engrev{
By contrast, the observations in November 2024, January 2025, and February 2025 yielded accuracies better than \rev{60~km} in Euclidean norm and a few tens of kilometers along the pulsar direction.
}
\engrev{
These results demonstrate that, although the accuracy of the position component along the pulsar direction remains consistently within approximately 40~km, the Euclidean norm accuracy depends substantially on the observation epoch.
}

\begin{table}[t]
\centering
\footnotesize               
\begin{threeparttable}[t]
\caption{\rev{
Values of orbital elements obtained via Bayesian optimization using the SEPO method for the four Crab Pulsar observation epochs.
}}
\label{tab:best_orbit}

\begin{tabular}{lcccccccccc}
\hline
Epoch
& \multicolumn{2}{c}{$B~(\times10^{-3}~R_{\oplus}^{-1})$}
& \multicolumn{2}{c}{$i~(^{\circ})$}
& \multicolumn{2}{c}{$\theta~(^{\circ})$}
& \multicolumn{2}{c}{$n-0.066~(\times10^{-5}~\mathrm{rad~min^{-1}})$}
& \multicolumn{2}{c}{$\Omega~(^{\circ})$} \\
& Opt.\tnote{a} & \rev{SD\tnote{b}} & Opt. & \rev{SD} & Opt. & \rev{SD} & Opt. & \rev{SD} & Opt. & \rev{SD} \\
\hline
2024 Apr & 0.66 &\rev{0.23}& 97.67 &\rev{0.26}& 357.19 &\rev{4.41}&  29.86 &\rev{0.24}& 167.62 &\rev{2.33} \\
2024 Nov & 2.07 &\rev{0.10}& 97.46 &\rev{0.09}& 360.12 &\rev{0.51}&  88.23 &\rev{0.21}& 21.28  &\rev{0.99} \\
2025 Jan & 1.46 &\rev{0.21}& 97.49 &\rev{0.75}& 360.14 &\rev{0.49}& 124.89 &\rev{0.36}& 103.80 &\rev{1.22} \\
2025 Feb & 1.70 &\rev{0.07}& 97.37 &\rev{0.36}& 360.58 &\rev{0.28}& 138.97 &\rev{0.14}& 134.59 &\rev{0.37} \\
\hline
\end{tabular}
\begin{tablenotes}
\footnotesize
\item[a]{\rev{Opt. means the best optimized value of orbital elements obtained from Bayesian optimization.}}
\item[b]{\rev{SD means the standard deviation of sampled values satisfying
$(\chi^2_{\rm max}-\chi^2_{\rm s})/\chi^2_{\rm max} < 0.005$.}}
\end{tablenotes}
\end{threeparttable}
\end{table}

\begin{table}[t]
\centering
\begin{threeparttable}[t]
\centering
\caption{Orbit determination accuracy for four Crab Pulsar observations using the SEPO method.
}
\label{tab:sepo_accuracy}
\begin{tabular}{lrrrr}
\hline
Epoch
& \multicolumn{2}{c}{Euclidian norm difference (km)}
& \multicolumn{2}{c}{Pulsar direction difference (km)} \\
& Opt.\tnote{a}&  \rev{SD}\tnote{b}& Opt.&  \rev{SD} \\
\hline
2024 Apr & \rev{3.7$\times10^2$} & \rev{3.1$\times10^2$} & \rev{34}& \rev{6} \\
2024 Nov &  \rev{27} & \rev{60} & \rev{13}& \rev{41}\\
2025 Jan &  \rev{49} & \rev{51} & \rev{18}& \rev{17}\\
2025 Feb &  \rev{53} & \rev{50} & \rev{27}& \rev{9} \\
\hline
\end{tabular}
\begin{tablenotes}
\footnotesize
\item[a]{\rev{
Opt. denotes the RMS position differences between the GPS-derived orbit and the SEPO-derived orbit obtained using the optimal orbital elements.
}}
\item[b]{\rev{
SD denotes the standard deviation of the RMS position differences between the GPS-derived orbit and orbits generated from sampled orbital element sets satisfying $(\chi^2_{\rm max}-\chi^2_{\rm s})/\chi^2_{\rm max} < 0.005$.
}}
\end{tablenotes}
\end{threeparttable}
\end{table}

\section{Discussion}
\label{sec:discussion}
Sheikh et al. proposed an analytical method for a simplified estimation of measurement precision from the pulse width of the pulsar and the signal-to-noise ratio (${\rm SNR}$) of the detected pulse~\cite{Sheikh2006-oj}. 
\engrev{If} the speed of light is denoted by $c~(=3.0\times10^8~{\rm m/s})$, the uncertainty of the angular measurement in the direction of the pulsar, $\sigma_{r}$, can be approximated using the \engrev{${\rm SNR}$ and FWHM} of the pulse width $W$ as
\begin{equation}
\sigma_{r} = c\frac{1}{2}\frac{W}{\rm SNR}.
\end{equation}
\engrev{
Let the total number of events be $N_{\rm total}$ and the number of pulsed signal events be $N_{\rm pulsed}$, ${\rm SNR}$ is denoted as $ N_{\rm pulsed}/\sqrt{N_{\rm total}}$.}
\rev{
$\sigma_{r}$ reflects the position error originating in the pulse shape and spin frequency of a pulsar and the statistical uncertainties of the observed X-ray photons.
}
\engrev{
The ${\rm SNRs}$ of observations in April 2024, November 2024, January 2025, and February 2025, calculated from the epoch-folded pulse profile of the Crab Pulsar barycentric corrected with ADCS orbit data, are respectively $1.6 \times 10^2$, $ 1.4\times 10^2$, $ 1.4\times 10^2$, and $1.6 \times 10^2$.
The numerical analysis with Nelson's formula shows that the $W$ of Crab pulse is 1.2~ms.
}
The estimated $\sigma_{r}$ of observations in April 2024, November 2024, January 2025, and February 2025 are \engrev{1.1, 1.3, 1.3, and 1.1~km in the direction of the Crab Pulsar, respectively}.
\rev{
These values are more than an order of magnitude smaller than the position differences presented in Table~\ref{tab:sepo_accuracy}.
}

As shown in Fig.~\ref{fig:TimingVerification_radio}(b), the deviation of the main phase offset from \engrev{the} JBO is within $0.3\times 10^{-2}$\engrev{, which corresponds to $1.0\times10^2$~$\mu$s in time, equivalent to an X-ray propagation distance of 30~km at $c$.}
\rev{
Table~\ref{tab:sepo_accuracy} shows that, even when the variation in pulse $\chi^2$ is below 0.5\%, the resulting position differences can be as large as 6--41~km.
Further improvements in positioning accuracy will require resolving distributions corresponding to smaller significance differences and reducing uncertainties associated with the optimization process.
The position differences relative to the GPS data in the Crab Pulsar direction are mainly attributable to uncertainties in the timing measurements and the Bayesian optimization procedure.
}

\engrev{Another} possible cause of the systematic deviation is that NinjaSat operates in an SSO, and observations are interrupted during passes through the auroral belts, the South Atlantic Anomaly, or Earth occultation. 
\engrev{Consequently}, only partial segments of the orbit contribute to the observational data used in the SEPO method analysis.
The optimization of orbital elements may have been affected by the exclusion of data from certain portions of the orbit, such as the polar region or the region opposite Crab Pulsar\engrev{, of the orbit}.

\engrev{We next discuss} the relationship between the sampled points in the Bayesian optimization, pulse significance, and the positioning accuracy, \engrev{considering} the geometrical relation between the orbital plane and the Crab Pulsar direction at each epoch.
The parameters, \engrev{$n$ and $B$}, correspond to the orbital period and its temporal variation, respectively, and therefore tend to be correlated in the Bayesian sampling.  
In the January 2025 dataset, where the observation span was relatively short, the significance became less sensitive to variations in the correlation between $n$ and $B$.

In April 2024, the separation angle between the orbital plane normal vector and \engrev{Crab Pulsar direction vector} was large, as shown in Fig.~\ref{fig:OrbPlaneAngYear}.  
Under this geometry, changes in the orbital phase at the start of the observation $\theta$ produced only small variations in the position along the Crab line of sight.  
\engrev{As} $\theta$ represents the orbital phase at the start of the observation and is directly related to the time information, the inability to track the short-term motion of the satellite \engrev{may significantly degrade accuracy}. 
\engrev{However}, variations in the inclination $i$ correspond directly to changes along the Crab line of sight, and thus\engrev{, the sampled points converge within a narrow range.}

In January 2025, when the separation angle between \engrev{the orbital-plane normal and the Crab Pulsar direction vectors} was small, as shown in Fig.~\ref{fig:OrbPlaneAngYear}, variations in $\theta$ were mapped into line-of-sight changes, resulting in a narrower distribution.  
\engrev{By} contrast, variations in $i$ and $\Omega$ produced only small line-of-sight changes, leading to a broader distribution of sampled points.  
Although the precision in $i$ and $\Omega$ was lost, the improved constraint on $\theta$ led to apparently \engrev{higher positioning accuracy than in} 2024 when evaluated against GPS data, as shown in Fig.~\ref{fig:ResultPosDiff}.  
However, the optimization of the orbital elements is unstable, and the result lacks robustness\engrev{, as shown in Fig.~\ref{fig:RegretPlot}.}

In November 2024 and February 2025, the orbital plane was inclined by \engrev{approximately} $45^\circ$ with respect to the Crab Pulsar direction.  
This geometry allowed variations in all orbital elements to be reflected as line-of-sight changes, providing the most stable optimization.
In summary, when performing orbital parameter searches with the SEPO method using a single pulsar, the variations in each orbital element are effectively projected onto the pulsar line of sight.  
Therefore, the most stable and accurate results are obtained in seasons when the variations \engrev{in} all orbital elements contribute to the line-of-sight motion.  
This demonstrates the importance of selecting appropriate pulsars based on the orbital dynamics and \engrev{observation epoch}.
In the November 2024 observations, at approximately $3\times10^2$~ks after the start of pulsar observations, and in the February 2025 observations, at approximately $4\times10^2$~ks, the position estimated from the pulsar signal surpassed that based on the initial orbital values, as shown in Fig.~\ref{fig:ResultPosDiff}.
\engrev{In observations where the geometrical relation between the orbit and pulsar direction was favorable, pulsar signals observed with the GMC are useful for semi-autonomous updates of orbital information.}
These results \engrev{show that, if} the information from TLEs or GPS signals becomes unavailable over time, the orbital information can be updated more accurately using pulsar observations with the GMC.

\section{Conclusion and Future Works}
\label{sec:conclusion}
\engrev{This study}  demonstrated the absolute timing verification and pulsar navigation capability of the CubeSat X-ray observatory NinjaSat using Crab Pulsar observation data. 
The onboard GMCs \engrev{measured} X-ray events timing with a resolution of 61~$\mu$s, and absolute timestamps were calibrated with GPS time data. 
\engrev{Relative to the JBO radio reference, the NinjaSat X-ray main peak was leading by 5.1$\times10^2$~$\mu$s, with a deviation of less than 100~$\mu$s throughout the observation.}
This 5.1$\times10^2$~$\mu$s \engrev{leading} corresponds to the expected offset between the radio and X-ray peaks of the Crab Pulsar.
In comparison with NICER in the 2--12~keV band, the average time difference was \rev{$8 \pm 1~\mu{\rm s}$} with a standard deviation of \rev{21~$\mu$s}.
Here, the error attached to the average represents the fitting error of the Gaussian peak in the cross-correlation analysis, and the standard deviation reflects the dispersion of the time offsets across \engrev{six} observations.

In the pulsar navigation experiments, we applied Bayesian optimization of orbital elements using Crab Pulsar data obtained in three different epochs.
\engrev{The navigation accuracy was strongly dependent on the angle between the orbital plane and pulsar direction.}
The positioning accuracy along the pulsar line of sight was consistently within \rev{$\sim$40~km}.
\engrev{When the pulsar direction was almost perpendicular to the orbital plane, short-term orbital motion could not be constrained, rendering absolute position determination challenging.}
Conversely, when the pulsar direction was \engrev{almost} parallel to the orbital plane, as in a sun-synchronous polar orbit, parameters that perturb the orbit in the normal direction became poorly constrained.
The most favorable condition was realized when variations in all orbital elements were effectively projected onto the pulsar line of sight, \engrev{enabling} the absolute position to be determined within a few tens of kilometers.

\engrev{For future practical applications, the number of candidate pulsars should be increased by improving detector sensitivity and by incorporating corrections for the binary orbital motion of neutron stars with companions and corrections for spin variations caused by mass accretion, thereby enabling adaptive pulsar selection depending on the orbital configuration.
In addition, the GMC clock was calibrated using GPS signals in this study; however, for deep-space applications, alternative methods will be required, as GPS data are unavailable.}
More autonomous timekeeping systems, such as onboard atomic clocks, will be required to enable navigation with the SEPO method independent of GPS.

\subsection*{Disclosures}
The authors declare that there are no financial interests, commercial affiliations, or other potential conflicts of interest that could have influenced the objectivity of this research or the writing of this paper.

\subsection* {Code, Data, and Materials Availability} 
The code and data used for this article are proprietary and are not publicly available. 
They are available from the corresponding author upon request.

\acknowledgments % equivalent to \section*{ACKNOWLEDGMENTS}   
Parts of this study have been reported in the SPIE conference proceedings~\cite{10.1117/12.3063833}, and the present article provides an extended and revised analysis based on those results.
This project was supported by JSPS KAKENHI (JP17K18776, JP18H04584, JP20H04743 JP24K00673, and JP25KJ0241).
N. O. was supported by RIKEN Junior Research Associate Program. 
T. E. was supported by “Extreme Natural Phenomena” RIKEN Hakubi project. 
A. A. and S.I. were supported by the RIKEN Student Researcher Program.
\rev{The authors thank the English editting service, Editage.}
The authors thank Chat GPT, Google Translate, Grammarly, Overleaf Toggle WriteFull, for language and grammar clean-up of this paper. 
The authors thank Dr. Yasuyuki Morita, Dr. Takahiro Nishi, Dr. Nobuyuki Kawai, and Dr. Matteo Bachetti for advice on data analysis.
This work was supported by the RIKEN Pioneering Projects.

\bibliography{report} 
Naoyuki Ota is a graduate student at Tokyo University of Science and a junior research associate at RIKEN.  
He received his BS and MS degrees in science from Tokyo University of Science in 2021 and 2023, respectively.
\bibliographystyle{spiebib} 
\end{spacing}
\end{document}